# Practical Solutions in Fully Homomorphic Encryption - A Survey Analyzing Existing Acceleration Methods

Yanwei Gong, Xiaolin Chang, Jelena Mišić, Vojislav B. Mišić, Jianhua Wang and Haoran Zhu

*Abstract*—Fully homomorphic encryption (FHE) has experienced significant development and continuous breakthroughs in theory, enabling its widespread application in various fields, like outsourcing computation and secure multi-party computing, in order to preserve privacy. Nonetheless, the application of FHE is constrained by its substantial computing overhead and storage cost. Researchers have proposed practical acceleration solutions to address these issues. This paper aims to provide a comprehensive survey for systematically comparing and analyzing the strengths and weaknesses of FHE acceleration schemes, which is currently lacking in the literature. The relevant researches conducted between 2019 and 2022 are investigated. We first provide a comprehensive summary of the latest research findings on accelerating FHE, aiming to offer valuable insights for researchers interested in FHE acceleration. Secondly, we classify existing acceleration schemes from algorithmic and hardware perspectives. We also propose evaluation metrics and conduct a detailed comparison of various methods. Finally, our study presents the future research directions of FHE acceleration, and also offers both guidance and support for practical application and theoretical research in this field.

## Content

| Nomenclature | |
|---|---|
| HE | Homomorphic Encryption |
| PHE | Partially Homomorphic Encryption |
| FHE | Fully Homomorphic Encryption |
| SWHE | Somewhat Homomorphic Encryption |
| HAdd | Homomorphic Addition |
| HMult | Homomorphic Multiplication |
| CUDA | Compute Unified Device Architecture |
| FU | Functional Unit |
| PE | Processing Element |
| CPU | Central Processing Unit |
| GPU | Graphics Processing Unit |
| FPGA | Field Programmable Gate Array |
| ASIC | Application Specific Integrated Circuit |
| PIM | Process in Memory |
| RNS | Residue number system |
| DSP | Digital Signal Processor |
| NTT | Number Theoretic Transforms |
| INTT | Inverse Number Theoretic Transforms |
| CRT | Chinese Remainder Theorem |
| DGT | Discrete Galois Transform |
| SIMD | Single Instruction Multiple Data |
| PRNG | Pseudo-Random Number Generator |

## 1 Introduction

The exponential-growing volume of data and the rapid development of cloud computing facilitate outsourced computation of big data [1][2]. However, the collected data contains a large amount of sensitive and private information, which can lead to privacy disclosure. There exist cryptographic technologies to safeguard privacy, among which is fully homomorphic encryption (FHE). FHE is being explored for protecting data privacy and then is applied to many application scenarios, especially those involving sensitive data, such as healthcare, finance and government [3]-[5]. It plays an important role in the field of privacy protection [6][7].

The concept of FHE was first proposed by Rivest *et al.* [8] in 1978, but the first FHE scheme was proposed by Gentry [14] in 2009. He also proposed a method for constructing an FHE scheme, which means that a partially homomorphic encryption (PHE) scheme can become an FHE scheme by using bootstrapping to add the noise refresh process. Based on this, researchers carried out a lot of researches on various methods of constructing an FHE scheme. The most representative schemes are BGV [15], FV [16], GSW [17], and CKKS [18]. In fact, FHE is just one of homomorphic encryption (HE), which also includes PHE [8]-[10] and somewhat homomorphic encryption (SWHE) [11]-[13]. FHE allows infinite calculation and supports both homomorphic addition (HAdd) and homomorphic multiplication (HMult) on the ciphertext. The key strength of FHE is that it offers cryptographically-strong privacy guarantees, but these guarantees come at the cost of massive computational overhead. Thus, many researchers have shifted their study attention to FHE acceleration. Although the FHE acceleration schemes have made good progress, there is still a lack of a summary of related works to support further development.

There are surveys about FHE. Moore *et al.* [19] gave a brief introduction to the existing FHE acceleration schemes, which only include those based on the graphics processing unit (GPU) and field programmable gate array (FPGA). They did not make an analysis about them, and there was no discussion of CKKS acceleration scheme because it was not proposed at that time. Acar *et al.* [20] introduced the development history of HE and concluded the theories and details of various typical algorithms of HE. At the same time, this paper also sorted out and compared the implementation libraries of some HE algorithms. Alaya *et al.* [21] summarized different HE application ways and provide the application scenarios, such as medical treatment, image, and other fields. Wood *et al.* [22] provided an overview of the application of FHE in medicine and bioinformatics, along with descriptions of how it can be realized by considering their different characteristics. Marcolla *et al.* [23] introduced the basic knowledge and security attributes of HE and further summarized the application scenarios of HE, such as machine learning, fog computing, and cloud computing. Moreover, it also introduced some libraries and tools for HE. However, these reviews neither referred to the FHE acceleration schemes nor provided a comprehensive summary of FHE acceleration schemes.

To bridge this gap, we collate existing studies and provide a comprehensive review of FHE acceleration schemes. In addition, we summarize and compare the related work in TABLE 1, which fully demonstrates the necessity of this survey. It is because no other survey has collated content similar to it.

TABLE 1
COMPARISON OF RELATED WORKS

| Ref | year | Theory summary | Scheme summary | Application summary | Acceleration summary | |
|---|---|---|---|---|---|---|
| | | | | | Algorithm optimization | Hardware optimization |
| [19] | 2014 | | | | | insufficient |
| [20] | 2018 | √ | √ | | | |
| [21] | 2020 | | | √ | | |
| [22] | 2020 | | | √ | | |
| [23] | 2022 | √ | | √ | | |
| Ours | 2023 | | | | √ | √ |

The main limitation of the application of FHE is its performance bottleneck, which means the huge cost of FHE can not satisfy the demand for practical application. As a result, a large number of studies began to study how to accelerate it. At present, there mainly exist two ways. On the one hand, it focuses on algorithm optimization to accelerate the FHE scheme itself. On the other hand, it uses hardware to accelerate it, such as Central Processing Unit (CPU), Graphics Processing Unit (GPU), Field Programmable Gate Array (FPGA), etc. Figure 1 displays the distribution of papers focusing on FHE acceleration across various hardware platforms during the last four years. Notably, a significant portion of the papers center on FHE acceleration via GPU and FPGA. It is speculated that while GPU may not represent the optimal hardware platform for FHE acceleration, GPU can be used relatively simply to accelerate FHE. Therefore, in the early stages of FHE acceleration research, it was more common for researchers to utilize GPU and achieve superior acceleration effects compared to using CPU. Figure 2 highlights the trend of papers on FHE acceleration based on different hardware platforms in recent years, thereby lending further support to the aforementioned speculation. The earlier stages of FHE acceleration research featured a greater emphasis on acceleration schemes based on GPU. However, later FPGA and ASIC were found to be more suitable hardware for FHE acceleration in terms of acceleration efficiency. This has resulted in an increasing number of papers on FHE acceleration based on them year after year.

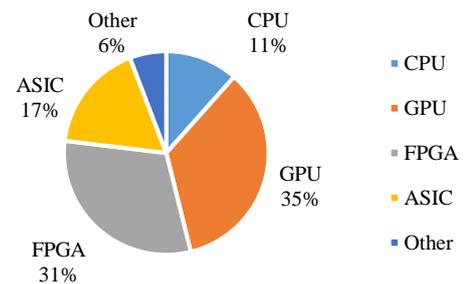

Figure 1. The distribution of papers focusing on FHE acceleration across various hardware platforms during the last four years

We review and summarize the current mainstream FHE acceleration schemes proposed in the last four years, with the following contributions:

- In this paper, we present a comprehensive summary and synthesis of the latest research findings on accelerating FHE between 2019 and 2022. The primary objective is to provide valuable insights for researchers interested in the state-of-the-art developments and future directions of FHE acceleration.

- Our study provides a comprehensive classification of existing acceleration methods for FHE from two perspectives: algorithmic acceleration and hardware acceleration. Further classification is conducted based on these two main categories, which is detailed in Section 3 and 4. Finally, we propose corresponding evaluation metrics to conduct a detailed comparison of various acceleration methods, aiming to offer guidance and inspiration for future research in this field. In particular, algorithmic-based methods achieve optimization by reducing the number of operations required for encryption, decryption, and homomorphic operations. Meanwhile, hardware-based methods focus on designing specialized hardware that can perform FHE operations more efficiently. Through this comprehensive comparison, we aim to provide a deeper understanding of the strengths and weaknesses of different FHE acceleration schemes.

- This paper explores the future research directions of FHE acceleration and provides guidance and support for practical application and theoretical research in this field. We identify various potential research directions, such as exploring new FHE algorithms, designing novel hardware architectures, and investigating hybrid acceleration schemes that combine algorithmic and hardware-based methods. By highlighting these potential research directions, we aim to encourage further advancements in FHE acceleration and its application to various fields for privacy preservation.

The rest of this paper is organized as follows. Section 2 introduces the basic knowledge to FHE acceleration schemes, including the knowledge related to algorithms and hardware. Section 3 and Section 4 respectively present the algorithm-based and hardware-based acceleration schemes. Following this, Section 5 gives the challenges and future research directions of FHE acceleration. Finally, Section 6 provides the conclusion. The organizational structure of this paper is shown in Figure 3.

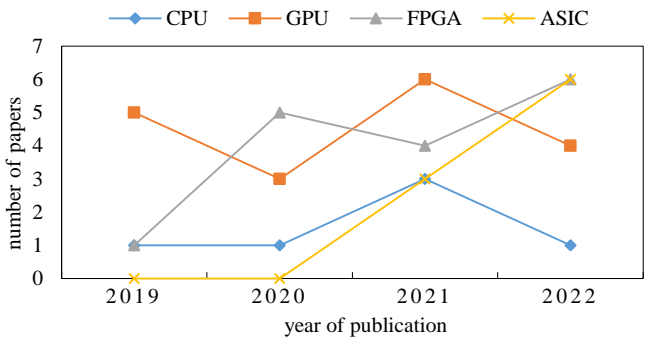

Figure 2. The trend of papers on FHE acceleration based on various hardware from 2019 to 2022

## 2 Preliminary

This section introduces the basic knowledge of FHE acceleration, including algorithms and hardware, in order to better understand FHE acceleration schemes discussed in Sections 3 and 4. Section 2.1 focuses on those operations related to FHE performance bottlenecks. Section 2.2 focuses on hardware platforms exploited to accelerate FHE schemes, including CPU, GPU, FPGA, and application specific integrated circuit (ASIC).

### 2.1 Algorithms about FHE

This section describes operations related to the performance bottlenecks of FHE, including polynomial addition, polynomial multiplication, number theoretic transform (NTT), and bootstrapping.

*1) Polynomial Multiplication*

Polynomial multiplication is an important homomorphic operation and the multiplication of very large degree polynomials is one of the major performance bottlenecks for the FHE implementations. NTT can be explored for FHE acceleration, which is done by first converting inputs to the NTT domain through polynomial multiplication. In the NTT domain, the polynomial operation can be converted to a coefficient-wise multiplication, which is also called dot multiplication. The coefficient-wise multiplication has high parallelism and is very suitable for executing on the hardware platform, which has a lot of parallel computing resources. The overhead of coefficient-wise multiplication is negligible when being compared to NTT and inverse NTT (INTT). They respectively represent the transformation of the polynomial multiplication input into the NTT domain and the reversal of the dot multiplication result, the overhead of coefficient-wise multiplication.

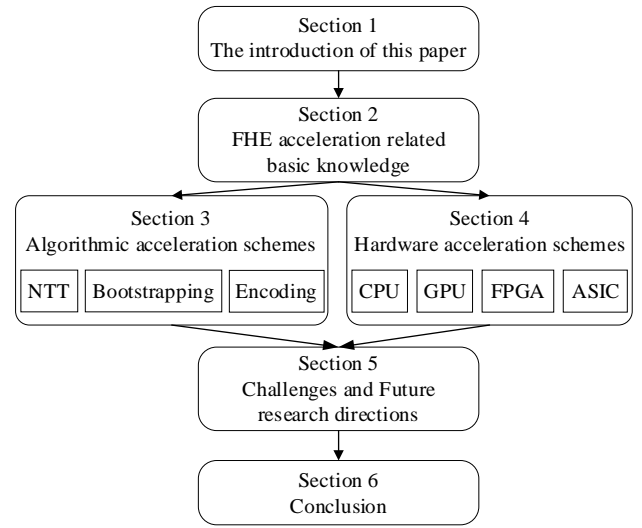

Figure 3. The organizational structure of this paper

*2) Polynomial Addition*

Like polynomial multiplication, polynomial addition is another important homomorphic operation in the homomorphic evaluation and is the second most frequently used operation after polynomial multiplication. Besides, the accumulation part of linearization also needs polynomial addition. The polynomial addition can be carried out in the Chinese remainder theorem (CRT) domain, which provides sufficient parallelism so that the hardware with parallel computing resources can be used to accelerate it.

*3) NTT*

NTT is extensively employed for FHE implementation because it enables fast polynomial multiplication by reducing the complexity of polynomial multiplication from $O(n^2)$ to $O(n \log n)$. NTT is defined as the discrete fourier transform over $\mathbb{Z}_q$. An $N$-point NTT operation transforms a polynomial $a$ into another degree polynomial $\tilde{a}$ with the same degree $N-1$. And the NTT can be naturally used for fast cyclic convolution.

*4) Bootstrapping Algorithm*

Bootstrapping algorithm is used to refresh the noise of ciphertext after a homomorphic evaluation when the entire noise budget of a ciphertext is consumed. It reduces the noise back to a lower level by running decryption homomorphically in order to allow an infinite number of computations of ciphertext. This is also the difference between FHE and SWHE. The bootstrapping operation consists of three major steps, including a linear transform, a polynomial evaluation, and another linear transform. All these steps consist of the same homomorphic operations, such as HAdd, HMult, and rotation.

*2.2 Hardware Platforms about FHE*

Section 2.2 describes the hardware used to accelerate FHE schemes. It analyses the characteristics of hardware in order to determine the optimal hardware platform for accelerating FHE schemes.

*1) CPU*

Compared with GPU, FPGA, and other hardware platforms, the acceleration effect achieved by CPU is not obvious. CPU itself is not designed to perform heavy data computing tasks. It can provide a degree of parallel computing power through multithreading, but this is far from enough to meet the needs of FHE acceleration. Therefore, most studies prefer to combine it with GPU or FPGA to realize the acceleration of FHE. However, CPU has the advantage of being a more general hardware platform with a wide range of applications.

*2) GPU*

The customization flexibility that GPU can provide is between CPU and FPGA. General-purpose computing on GPU yields greater efficiency when standardized by price compared to FPGA and ASIC. GPU is a powerful but highly specialized device that requires careful coding to take full advantage of the large amount of parallelism it offers. Specifically, the programming model and memory organization are quite different from the CPU. Compared with FPGA, it is easier to develop programs. In addition, there is a compute unified device architecture (CUDA) toolkit to facilitate program development.

*3) FPGA*

FPGA is a chip that can be reconfigurable circuitry. It is a hardware reconfigurable architecture, so it can provide more customization flexibility. Both CPU and GPU belong to the von Neumann structure, while FPGA is a no-instruction and no shared memory architecture. This structure makes FPGA much more energy efficient than CPU or even GPU. The function of each logic unit in an FPGA is determined during reprogramming, without the need for instructions. FPGA has both pipeline and data parallelism, whereas GPU only has data parallelism (pipeline depth is limited). Due to the flexibility of FPGA, more researchers focus on the acceleration of FHE using FPGA.

*4) ASIC*

ASIC is capable of achieving maximum acceleration effects concerning FHE, while it requires customization and therefore their application scope is limited. The biggest advantage of ASIC is that it can be designed corresponding circuits completely according to the needs of computing tasks, thus providing maximum acceleration capability. However, ASIC needs to go through many processes from being designed to be put into use, and it is also costly. So its application in FHE acceleration is more in the simulation stage. But the idea of using ASIC to accelerate FHE can be borrowed and applied to other hardware platforms.

Each FHE accelerator has its own strengths and weakness. CPU is the most common computer processor and can be used to implement FHE schemes. However, due to its single-instruction pipeline architecture, its parallel performance and processing speed are relatively low, making it less suitable for high-performance homomorphic operations. GPU has excellent parallel computing capabilities when processing large-scale data, so they can be used to accelerate FHE schemes. However, since GPU are designed for graphics processing, their support for numerical computing is limited, and sometimes it is difficult to adapt to the special requirements of FHE schemes. FPGA is a programmable logic device with high flexibility and parallel performance, and can be used to implement high-performance FHE schemes. Since FPGA is programmable, it can be customized and optimized according to specific needs and application scenarios, thereby achieving higher performance and efficiency. ASIC is a chip specially designed and manufactured to implement highly customized and optimized FHE schemes. Compared with FPGA, ASIC has higher performance and lower power consumption, but also has higher design costs and longer development cycles.

In summary, the application scenario and specific requirements must be considered for selecting an accelerator in order to achieve the optimal performance and efficiency.

TABLE 2
COMPARISON OF HARDWARE PLATFORMS

| Hardware platform / Metric | CPU | GPU | FPGA | ASIC |
|---|---|---|---|---|
| Universality | +++++ | ++++ | +++ | ++ |
| Customization flexibility | + | +++ | ++++ | +++++ |
| Acceleration ability | + | +++ | ++++ | +++++ |
| Price | ++++ | +++ | ++ | +++++ |
| Practicality | + | ++++ | ++++ | +++ |
| Research popularity | + | ++++ | ++++ | +++ |

\* Note that We use the number of '+' to describe the strength of hardware with respect to a feature, with five '+' representing the strongest and one '+' representing the weakest. The same is true for the meaning of '+' in the other tables presented by this paper.

TABLE 2 compares the hardware platforms from 6 aspects. *Universality* refers to whether the hardware platform is widely used, and it is persuasive to say that CPU is the most widely used hardware platform. Whether or not a hardware platform provides enough customization can be evaluated through customization flexibility, which also reflects the acceleration ability it can offer. ASIC is considered to be able to provide the largest customization flexibility because it allows you to design hardware circuits. As far as practicality is concerned, there are two aspects to consider: ease of use and acceleration ability. In short, if a hardware platform provides good acceleration but with low usability, it would be regarded as having low practicality, such as ASIC. The metric of research popularity, same as in TABLE 3, is determined according to the number of related papers.

## 3 Algorithm Acceleration Schemes

This section introduces the acceleration schemes of FHE using algorithm optimization, which is mainly divided into three categories, including NTT, Bootstrapping, and Encoding. TABLE 3 compares the characteristics of them, including acceleration approaches, importance, acceleration effect, and research popularity. It can be seen that NTT and bootstrapping are studied by more researchers. TABLE 4 gives all the acceleration schemes discussed in Section 3 and 4, and also shows whether they are algorithm-based.

### 3.1 NTT Optimization

As a very important primitive operation in FHE, NTT has very important research value for FHE acceleration. Therefore, a large amount of research works focus on the design of acceleration schemes for NTT, including algorithm optimization and hardware optimization. The core idea of NTT algorithm optimization is to use the existing algorithm to simplify the NTT operation and replace it with a more suitable form of hardware parallel computing.

Agrawal *et al.* [52] accelerate FHE by using algorithmic optimization. The most important part of this paper in terms of algorithmic optimization is to use low-cost operations to replace high-cost operations. Barrett reduction [88] method is adopted to transform the high-consumption modular operations. In addition, the NTT operation in polynomial multiplication is used to perform the indices computation operation by shifting and XOR operation, so as to accelerate the NTT process. Similar to [52], Badawi *et al.* [45] also use Barrett Reduction to accelerate the implementation of the algorithm.

Compared to [52], not only do Shivdikar *et al.* [46] use Barrett reduction, but they also improve it. Based on several variants of Barrett reduction, an efficient Barrett reduction for 64-bit integers is proposed to accelerate the modular division of NTT operations used in polynomial multiplication operations. Mert *et al.* [49] accelerate the encryption and decryption process of BFV algorithm. Unlike [46][52], an efficient polynomial multiplier is proposed which can also be used for homomorphic operations other than encryption and decryption. They mainly use Montgomery algorithm to reduce the modular division operation in polynomial multiplication, so as to achieve the algorithm acceleration. Goey *et al.* [40] not only use Barrett reduction but also use SSMA algorithm [85] to further accelerate NTT operation. SSMA is a fast multiplication algorithm for large integers with a low computational complexity of $O(n \log(n) \log(n \log(n)))$. Therefore, in terms of acceleration results, cuHE [86] is surpassed. Roy *et al.* [48] accelerate the BFV. But they only use the existing latest algorithm level optimization method.

The algorithm acceleration for NTT mainly focuses on utilizing Barrett reduction. Some studies [45][52] just use it, while other [46] improves it. Moreover, some studies [40][49] also use other fast multiplication algorithms to accelerate NTT.

### 3.2 Bootstrapping Optimization

Bootstrapping, which has huge time complexity and space complexity, is another important performance bottleneck in FHE. Therefore, the core idea of bootstrapping optimization is to solve the memory bandwidth problem in its implementation process and improve throughput.

Chen *et al.* [24] focus on bootstrapping [97] acceleration of CKKS. They use a dynamic programming approach [93] to decide the optimal level collapsing strategy for a generic multi-leveled linear transform in order to fully explore the trade-off between levels consumption and the number of operations. And the result shows a large increase in the bootstrapping throughput. In addition, they replace the Taylor approximation with the Chebyshev interpolant to approximate the scaled sine function, which not only consumes fewer levels but also is more accurate than the original method.

For the full-residue number system (full-RNS) variant of CKKS, Han *et al.* [25] combine the RNS-decomposition method [95] and the temporary modulus technique [96] to reduce about half complexity for HMult even with a larger security parameter. For the evaluation of *sine* function and *cosine* function, they consider a ratio between the size of a message and the size of a ciphertext modulus. As a result, the number of non-scalar multiplications is almost reduced by half compared to the previous work [24].

Bossuat *et al.* [26] accelerate the bootstrapping for the full-RNS variant of the CKKS. They propose a new format for rotation keys and a modified key-switching procedure in order to improve the baby-step giant-step algorithm [98], which is used by previous works [24][25][99]. The modified key-switching procedure extends the hoisting [76] technique to a second layer and reduces the cost of the linear transformations compared to the previous hoisting approach. Moreover, they also discuss the parametrization of the CKKS and its bootstrapping circuit and propose a procedure to choose and fine-tune the parameters for a given use-case.

TABLE 3
COMPARISON OF ACCELERATION OBJECT

| Accelerated object | Acceleration approaches | Importance | Research popularity | Accelerated effect |
| --- | --- | --- | --- | --- |
| NTT | Operational substitution exploration<br>Algorithm parallelism exploration | +++++ | +++++ | ++++ |
| Bootstrapping | Data dependency exploration<br>Algorithm simplification exploration | +++++ | +++ | +++ |
| Encoding | Efficient coding based on the data access mode of the subsequent algorithm | + | + | + |

TABLE 4
COMPARISON OF DIFFERENT ACCELERATION SCHEMES

| Ref | Year | Algorithm-based optimization | | | Hardware-based optimization | | | | |
|---|---|---|---|---|---|---|---|---|---|
| | | NTT | Bootstrapping | Encoding | CPU | GPU | FPGA | ASIC | Other |
| [24] | 2019 | √ | | | √ | | | | |
| [25] | 2020 | √ | | | √ | | | | |
| [26] | 2021 | √ | | | √ | | | | |
| [27] | 2021 | | | | √ | | | | |
| [28] | 2021 | | | | √ | | | | |
| [29] | 2022 | | | | √ | | | | |
| [30] | 2019 | | | √ | | √ | | | |
| [31] | 2019 | | | | | √ | | | |
| [32] | 2019 | | | | | √ | | | |
| [33] | 2019 | | | | | √ | | | |
| [34] | 2019 | | | | | √ | | | |
| [35] | 2020 | | | | | √ | | | |
| [36] | 2020 | | | | | √ | | | |
| [37] | 2020 | | | | | √ | | | |
| [38] | 2021 | | | | | √ | | | |
| [39] | 2021 | | | | | √ | | | |
| [40] | 2021 | √ | | | | √ | | | |
| [41] | 2021 | | | | | √ | | | |
| [42] | 2021 | | | | | √ | | | |
| [43] | 2021 | | √ | | | √ | | | |
| [44] | 2022 | | | | | √ | | | |
| [45] | 2022 | √ | | | | √ | | | |
| [46] | 2022 | √ | | | | √ | | | |
| [47] | 2022 | | | | | √ | | | |
| [48] | 2019 | √ | | | | | √ | | |
| [49] | 2020 | √ | | | | | √ | | |
| [50] | 2020 | √ | | | | | √ | | |
| [51] | 2020 | | | | | | √ | | |
| [52] | 2020 | √ | | | | | √ | | |
| [53] | 2020 | | | | | | √ | | |
| [54] | 2021 | | | | | | √ | | |
| [55] | 2021 | | | | | | √ | | |
| [56] | 2021 | | √ | | | | √ | | |
| [57] | 2021 | | | | | | √ | | |
| [58] | 2022 | | | | | | √ | | |
| [59] | 2022 | | | | | | √ | | |
| [60] | 2022 | | | | | | √ | | |
| [61] | 2022 | | | | | | √ | | |
| [62] | 2022 | | | | | | √ | | |
| [63] | 2022 | | | | | | √ | | |
| [64] | 2021 | | | | | | | √ | |
| [65] | 2021 | | | | | | | √ | |
| [66] | 2021 | | | | | | | √ | |
| [67] | 2022 | | | | | | | √ | |
| [68] | 2022 | | √ | | | | | √ | |
| [69] | 2022 | | | | | | | √ | |
| [70] | 2022 | | | | | | | √ | |
| [71] | 2022 | | | | | | | √ | |
| [72] | 2022 | | | | | | | √ | |
| [73] | 2021 | | | | | | | | √ |
| [74] | 2022 | | | | | | | | √ |
| [75] | 2022 | | | | | | | | √ |

*Note that the article with a '√' for both algorithm-based optimization and hardware-based optimization, represents that it leverages both algorithm optimization and hardware optimization simultaneously.

In [43], the optimization method of bootstrapping is proposed aiming at improving the throughput. The authors present algorithmic optimizations including combining ModDown and rescale in Mult, hoisting the ModDown in PtMatVecMult, and compressing the key with a pseudo-random number generator (PRNG). By combining ModDown and rescale in Mult, they realize a faster encrypted inner product. The ModDown in PtMatVecMult leads that the same ciphertext can be computed more efficiently than simply applying the rotate function. Compressing the key with a PRNG can avoid shipping the large random polynomials to dynamic random access memory, instead sending only the short PRNG key. The three optimization methods can reduce the number of operations and the consumption of memory, thus realizing the acceleration of bootstrapping.

Ye *et al.* [56] accelerate the application of FHE in CNN. They mainly use a new convolution method and explore the parallelism of the algorithm to realize the algorithm acceleration. By using Im2col convolution and Frequency domain convolution, they guarantee that any pair of elements to be summed or multiplied are in the same position in the vector. Therefore, only Pt-Ct Mult and Additions without the expensive rotations are needed for homomorphic convolution calculations.

Kim *et al.* [68] propose the Minimum key-switching (Min-KS) based on the minimal key-switching proposed by Halevi *et al.* [76]. Compared with minimal key-switching [76], Min-KS further reduces the use of the evaluation key in bootstrapping, which means that the number of data access in the chip memory at a time is decreased. In addition, the Min-KS algorithm is generalized so that it can be better applied to common homomorphic operations.

Since bootstrapping involves many primitive operations, many studies have been done to accelerate it by speeding up some of them. The key-switching procedure is the focus of some research works [26][43][56][68], which reduce the number of operations to achieve the purpose of acceleration.

### 3.3 Encoding Optimization

The core idea of encoding optimization is to make the encoded data have more parallel computing potential, so as to accelerate the implementation of the algorithm. However, there are a few research works on algorithm acceleration by using encoding optimization.

Jin *et al.* [30] design an encoding strategy for FHE, so that the encoded plaintext data can be better executed in parallel. Compared with the previous work, they design the corresponding data encoding strategy for images with higher dimensions. In view of the feature that different weight components are calculated simultaneously in convolution operation, they encode the components that can be parallel into a vector, so as to facilitate the parallel homomorphic operation later. Besides, the encoding strategy improves memory efficiency and reduces the message size transferred.

### 3.4 Algorithm Acceleration Summary

In general, the acceleration effect of FHE schemes based on algorithm acceleration is limited. Many research works on the acceleration of FHE focus on the acceleration of NTT and bootstrapping. And the Barrett reduction is most commonly used to accelerate NTT. There are two main ways of algorithm optimization. One is to replace operations, which needs high computing resource consumption, with operations consuming high computing resource. The other is to improve the parallelism of the algorithm so that the hardware acceleration schemes can be designed on this basis. In general, the algorithm acceleration schemes are unable to achieve a breakthrough acceleration effect and most works just apply existing algorithms. Therefore, algorithm-based acceleration schemes require a theoretical breakthrough, especially for bootstrapping, in order to make FHE performance closer to the actual application requirements. However, the advantage of an acceleration scheme based on algorithm optimization is that it is more adaptive and can adapt to different hardware platforms.

## 4 Hardware Acceleration Schemes

This section introduces FHE acceleration schemes using hardware platforms, including CPU, GPU, FPGA, ASIC, and Other. CPU has the least parallel computing resources, resulting in the worst acceleration effect. GPU is cheaper and easier to use. FPGA is a popular choice for FHE acceleration due to its parallel computing resources and customization. ASIC achieves the best acceleration effect although it's difficult to put into production.

### 4.1 CPU-based

There are only a few CPU-based acceleration schemes, and the acceleration effect that can be achieved is limited.

Based on the Intel® Advanced Vector Extensions 512 (Intel® AVX512) instruction set, Boemer *et al.* [27] implement acceleration on polynomial modular multiplication and NTT. They provide it as a Homomorphic Encryption Acceleration Library which can be used in combination with the SEAL library. The acceleration scheme mainly uses two optimization methods, including loop optimization and data parallel computing optimization, which are both based on the instruction set. Loops are unrolled either manually or using a pre-processor directive, with a manually-tuned unrolling factor. Within manually unrolled loops, instructions are reordered where possible for best pipelining. Based on the property of the instruction set, eight 64-bit integers can be processed simultaneously, the acceleration scheme completes the acceleration of NTT by expressing NTT and INTT in the form of elements and processing them together. For the vector and vector multiplication and polynomial and polynomial multiplication expressed in the form of elements, the input data is also aligned based on the purpose for simultaneous processing of eight 64-bit integers to achieve performance optimization.

In Ishimaki *et al.* [28], the Trace-Type Function Evaluation of CKKS is accelerated. The homomorphic trace-type function evaluation is performed by repeating homomorphic rotation followed by addition (rotations-and-sums). The homomorphic trace-type function is a commonly used and time-consuming subroutine that enables homomorphically summing up the components of the vector or homomorphically extracting the coefficients of the polynomial. They propose a more efficient trace-type function evaluation using loop-unrolling to reduce the number of expensive operations by leveraging a property of automorphisms and using a multi-core environment, thus reducing the computational cost compared with the sequential method [93][94] at the expense of slightly increasing the required storage. In addition, they successively unroll consecutive subloops in the trace-type function evaluation and parameterize the number of iterations after the unrolling, which further realizes the acceleration of the Homomorphic Trace-Type Function Evaluation.

Inoue *et al.* [29] also accelerate the trace-type function by using Intel® AVX512 instruction set. Through using loop unrolling, they implement the optimization of the trace-type function. Their acceleration scheme can be regarded as another implementation scheme of Intel® AVX512 instruction set in accelerating FHE, without much innovation.

Most studies [27][29] are based on the Intel® AVX512 instruction set to accelerate FHE, since the mainstream CPU instruction set is not suitable. This also shows the limitations of the CPU for FHE acceleration. TABLE 5 shows the comparison of acceleration schemes based on CPU. In it, three aspects should be considered in the availability evaluation of schemes. They are the acceleration ability offered, whether the hardware platforms are universal, and whether the schemes are only accelerated architectures or form acceleration libraries that could be directly called.

### 4.2 GPU-based

GPU has been widely used in FHE acceleration in recent years. The acceleration scheme based on GPU is mainly realized by utilizing parallel computing resources it provides. At the same time, the data storage strategy will also be optimized in order to reduce the time taken to access data. However, due to the limited customization flexibility provided by GPU, the innovation of GPU-based acceleration schemes is also limited. On the other hand, since GPU is a more general hardware platform, its acceleration scheme can be presented in the form of the acceleration library and put into use.

Badawi *et al.* [31] implement and evaluate the performance of two optimized variants, which are called Bajard Eynard-Hasan-Zucca (BEHZ) and Halevi-Polyakov-Shoup (HPS), based on CPU and GPU. The lazy reduction and several precomputations are added to optimize the implementation of HPS. When implementing HPS on GPU platform, the discrete galois transform (DGT) is used for efficient polynomial multiplication using negacyclic convolution. It cuts the transform length into half and requires less amount of memory for precomputed twiddle factors. The DGT algorithm was originally proposed by Crandall [80] for fast negacyclic convolution. Besides, data transfer between CPU and GPU is avoided. The final result shows that HPS performs better than BEHZ, and for 128-bit security settings, HPS is already practical for cloud environments supporting GPU computations.

TABLE 5
COMPARISON OF ACCELERATION SCHEMES BASED ON CPU

| Ref | Year | Availability | Optimize data storage policy | Homomorphic parameters tuning | Accelerated object |
|---|---|---|---|---|---|
| [27] | 2021 | + | √ | × | Bootstrapping |
| [28] | 2021 | + | √ | √ | Rotation / HAdd |
| [29] | 2022 | + | √ | × | Rotation / HAdd |

Lupascu *et al.* [32] use several GPUs to accelerate the FHE and first adapt them for HElib. The acceleration scheme proposed in this paper is to make full use of parallel computing resources of multiple GPUs by distributing computing tasks reasonably. For task allocation on multiple GPUs, the unity of task loads is balanced to maximize the work efficiency of multiple GPUs. Before tasks are distributed, for the ciphertext addition, subtraction, and multiplication operations in homomorphic operations, the parallelization of the operation sequence is explored, and the operation sequence that can be executed in parallel is decomposed. The innovation of the acceleration scheme proposed in this paper is limited, and it is just to accelerate the algorithm implementation at the cost of consuming multiple GPUs.

Similar to [32], Lei *et al.* [33] also use GPUs to accelerate the adder in FHEW-V2. The cuFTT [81] library is used to accelerate the parallelization of FTTs operations in bootstrapping. For a complex multiplication operation, they use the parallelism of GPU to accelerate it. In addition, for the data that need to be shared in the calculation, they put it into the shared memory to improve the data access speed. Different from [32], their acceleration scheme also uses a multicore CPU to accelerate the algorithm. For example, since bootstrapping is independent of each key, this paper uses a multicore CPU to accelerate the key generation process by generating different keys at the same time.

Based on the joint architecture of CPU and GPU, Xia *et al.* [34] accelerate the DGHV [82]. Aiming at serialization operations of DGHV, they explore the method of parallel implementation and use parallel computing resources of GPU to carry out the implementation of corresponding algorithms. For example, for the long message sequence, the acceleration scheme adopts the method of dividing the message into blocks and then simultaneously encrypting each message block. For HAdd, the acceleration scheme also uses the computing resources of GPU to implement it in parallel. In general, the acceleration scheme [34] is relatively simple, and the exploration of algorithm parallelism is not very sufficient. So the final acceleration effect is limited.

Different from studies mentioned above, based on GPU, Kim *et al.* [35] aim at the severe memory bandwidth bottleneck faced by NTT operation under a larger homomorphic parameters set. An implementation scheme of runtime data generation is proposed, which uses factorization recursion to calculate the rotation factor, thus balancing the calculation speed and space consumption.

Based on CPU and GPU, Morshed *et al.* [36] design and implement an acceleration scheme for TFHE homomorphic encryption. On the one hand, they analyze the parallelism for HAdd, HMult, and other operations, and realize the algorithm acceleration by assigning tasks, which can be computed in parallel, to different threads provided by CPU. On the other hand, they improve the parallelization of the algorithm by optimizing Bit Coalescing, Compound Gate, Karatsuba Multiplication [83], and so on, so as to make full use of the parallel computing resources of GPU.

Badaw *et al.* [37] realize the acceleration of CKKS based on GPU, while the important optimization they used here is related to matrix packing. Packing is useful in FHE to reduce both the number of homomorphic operations due to single instruction multiple data (SIMD) evaluation and the number of ciphertexts. In general, the innovation of the acceleration scheme proposed in this paper is limited, and it can not achieve an obvious acceleration effect.

Based on multiple GPUs, Badawi *et al.* [38] also design a data allocation strategy for the computation process of the FHE algorithm, which ensures the load balancing of multiple GPUs. According to the size of the polynomial matrix, the strategy allocates data with limited rows and columns, which achieves acceleration similar to the task allocation strategy proposed by Lupascu *et al.* [32]. However, Badawi *et al.* [38] also design an efficient CPU-GPU communication protocol, thus realizing better algorithm acceleration.

In Alves *et al.* [39], the acceleration scheme of BFV algorithm is designed based on GPU. Due to the inspiration of Bailey's version of the Fast Fourier Transform [92], they proposed a novel hierarchical formulation of the DGT that offers about two times lower latency than the best version available in the previous work. The DGT is also faster than NTT due to its lower memory bandwidth requirement. Moreover, they also choose the compatible parameters between the DGT and the RNS representation. In addition, a more efficient and polynomial-oriented state machine is designed to reduce the need for moving data in and out of the DGT domain and between the main memory and the GPU global memory. In a word, in order to better play the performance of the GPU, this paper innovatively uses some mathematical tools and implements them on the GPU through CUDA.

Goey *et al.* [40] accelerate CMNT [84] based on GPU. In terms of hardware optimization, they propose to buffer the twiddle factors in GPU registers and then access these data across different threads during the NTT computation through warp shuffle instruction. Combined with algorithm optimization mentioned in Section 3.1, the acceleration scheme finally realizes the further surpassing of cuHE.

Jung *et al.* [41] explore the parallel implementation and data access of CKKS. Based on this, the corresponding acceleration schemes are implemented on CPU and GPU and the result shows the effectiveness of the acceleration method. For CPU, the acceleration scheme is based on scatter instructions in AVX-512 [27], which accelerates matrix multiplication. For GPU, they assign each independently computable output element to a thread, so as to realize the acceleration of the algorithm through parallel calculation of non-interference data. This also makes the acceleration scheme more sufficiently explore the parallelism of algorithm implementation compared with cuHE. In addition to the design of parallel strategy, they also calculate and store data in advance to speed up data access.

Jung *et al.* [42] implement the optimization of CKKS bootstrapping. By fully exploring the parallelism in FHE, they discover that the major performance bottleneck is their high main-memory bandwidth requirement, which is exacerbated by leveraging existing optimizations targeted to reduce the required computation. Further, they find that inner product in key-switching is a major bottleneck when attempting to accelerate HMult. On this basis, kernel fusion and reordering primary functions, which are memory-centric optimizations, are used to accelerate bootstrapping. And the result shows that the acceleration scheme has achieved good acceleration effect.

Through the analysis of bootstrapping, Castro *et al.* [43] find that the main performance bottleneck of bootstrapping mainly comes from the memory bandwidth bottleneck. That is, a large number of intermediate process data needs to be generated and stored during the execution of bootstrapping. On this basis, the optimization method of bootstrapping is proposed, which mainly uses physical address mapping optimization. The physical address mapping of the data in the main memory has a substantial impact on the time it takes to transfer data. The reduction of data transfer time speeds up the implementation of bootstrapping.

Based on GPU, Özerk *et al.* [44] design and implement a corresponding acceleration scheme for key generation, encryption, and decryption of FHE using NTT operation. In this paper, the parallel operation of NTT is designed based on the parallel computing resources provided by GPU. At the same time, they also optimize the GPU memory usage and kernel function call, so as to realize the algorithm acceleration. The difference is that they focus on different acceleration objects mentioned above.

Türkoğlu *et al.* [45] design and implement a corresponding acceleration scheme for HMult, relinearization, rotation, and HAdd. They use the CUDA to further accelerate the Barrett reduction process. And they provide the corresponding GPU FHE acceleration library.

Based on GPU, Shivdikar *et al.* [46] design an acceleration method for polynomial multiplication, which is a very important performance bottleneck in homomorphic encryption. The method optimizes the memory access of GPU to improve the data access speed and throughput during the algorithm operation process, so as to achieve further acceleration.

The acceleration scheme proposed by Shen *et al.* [47] is GPU-based and accelerated for word-size FHE, including BGV, BFV, and CKKS. They mainly combine some previous acceleration optimization methods and develop the acceleration library on this basis. This acceleration library can be used for both ordinary GPU and embedded GPU. Meanwhile, it can better support the use of Internet of Things devices. And aiming at the security vulnerabilities of FHE acceleration stock in preventing side-channel attacks, they implement time consistency for multiply-accumulation, conditional subtraction, and Barrett reduction. At the same time, it also carried on the simplification of the implementation instructions. Besides, they also implement two versions for NTT operations, which are performance first and memory first. In view of the limited performance, the operation is further accelerated by simplifying the instructions. For polynomial multiplication operations and ciphertext multiplication operations, a general RNS multiplication kernel is designed to speed up the calculation. In a word, compared with the FHE accelerated GPU library [45], the acceleration library has three advantages, which are better adaption, higher security, and more features.

Due to the need for CPU cooperation, most studies [31][34][36][38][41] based on GPU acceleration require support from CPU. However, schemes designed by these studies focus on using GPU to accelerate FHE. Thus, we have classified these schemes as GPU-based FHE acceleration schemes. Research on GPU-based FHE acceleration started earlier, and currently, GPU-based acceleration schemes are gradually transitioning from proposing acceleration architectures to form acceleration libraries [45][47], which greatly improves availability. Additionally, some works [32][33] utilize multiple GPUs instead of a single GPU for accelerating FHE. TABLE 6 shows the comparison of acceleration schemes based on GPU. We can see that all FHE acceleration schemes based on GPU do not utilize homomorphic parameters tuning. This may be because parameters tuning itself is a complex process, and early acceleration of FHE could achieve good results by just using GPUs. And researchers have not delved deeply into the acceleration of GPU-based schemes. Hence, incorporating homomorphic parameters tuning is also a direction for research into accelerating FHE using GPU.

*4.3 FPGA-based*

Compared with GPU, FPGA can provide more flexible parallel computing resources. Therefore, the acceleration schemes based on FPGA will further explore the parallel computing implementation of FHE. In addition, the data access patterns and data placement strategies are also the focus of researchers. Reasonable data placement can not only speed up data access but also improve the parallelism of algorithm execution. In short, the acceleration scheme based on FPGA can achieve a better acceleration effect compared with GPU on the whole. However, the corresponding acceleration scheme is also more difficult to design.

TABLE 6
COMPARISON OF ACCELERATION SCHEMES BASED ON GPU

| Ref | Year | Availability | Optimize data storage policy | Homomorphic parameters tuning | Accelerated object |
|---|---|---|---|---|---|
| [31] | 2019 | + | × | × | HPS RNS variant of the BFV |
| [32] | 2019 | +++ | × | × | Ciphertext multiplication |
| [33] | 2019 | +++ | × | × | KeyGen / Bootstrapping |
| [34] | 2019 | ++ | × | × | DGHV |
| [35] | 2020 | +++ | √ | × | NTT |
| [36] | 2020 | ++ | × | × | TFHE |
| [37] | 2020 | + | × | × | CKKS |
| [38] | 2020 | ++ | × | × | HPS RNS variant of the BFV |
| [39] | 2021 | +++ | × | × | BFV |
| [40] | 2021 | ++ | √ | × | CMNT |
| [41] | 2021 | ++ | √ | × | CKKS |
| [42] | 2021 | ++++ | √ | × | Bootstrapping |
| [43] | 2021 | + | × | × | NTT / Polynomial Multiplication |
| [44] | 2022 | +++ | √ | × | NTT / Polynomial Multiplication |
| [45] | 2022 | +++ | × | × | HMult / Relinearization / Rotation / HAdd |
| [46] | 2022 | ++++ | √ | × | Polynomial Multiplication |
| [47] | 2022 | ++++ | × | × | BGV / BFV / CKKS |

Based on the joint architecture of ARM-FPGA, Roy *et al.* [48] accelerate the BFV. They mainly use the hardware parallel computing resources to design the hardware acceleration architecture and finally realize it, which is more efficient than the software implementation. The acceleration scheme is based on the NTT parallel computing method [77] and at the same time they make improvements to it, which is that on-chip memory is used to store a constant rotation factor to save cycles for further acceleration.

Mert *et al.* [49] accelerate the encryption and decryption process of BFV by using FPGA. In this paper, a more efficient parallel hardware architecture is proposed for NTT operation, which is divided into two parts. Then the input of each part is calculated in parallel, and finally the calculation results are combined to complete the NTT operation, thus realizing the acceleration of NTT operation. In addition, the most important part of this acceleration scheme is to implement a faster polynomial multiplier using FPGA. And compared with the acceleration scheme [78], the acceleration scheme [49] can adapt to larger homomorphic parameters.

Riazi et al [50] (HEAX) are based on FPGA to optimize the performance of CKKS, which explores multi-level parallelism for the implementation of algorithms and gives the corresponding optimization method and implementation framework. For HMult, HEAX first modifies the SEAL library so that more parameters can be adapted. In addition, it realizes optimization of polynomial multiplication by storing multiple correlation coefficients of a polynomial in memory space that can be accessed in parallel, thus speeding up the reading and writing speed of the data access. At the same time, the data access mode in NTT process is analyzed. In view of the key-switching operation, the pipeline architecture is designed by analyzing the data dependency, which can execute many key-switching operations simultaneously. As can be seen from the above introduction, the biggest advantage of HEAX is that it sufficiently explores the parallelism of algorithm implementation and carries out optimization implementation on this basis. Therefore, a good acceleration effect has been achieved by HEAX.

Kim *et al.* [51], based on FPGA, mainly accelerate NTT operation. Compared with acceleration schemes [48][50], they focus more on bootstrapping and RNS domain and present a novel hardware architecture of NTT. In the design of the acceleration scheme, they take into account the algorithm to realize the time consumption and space consumption, which maximizes the use of resources. Aiming at butterfly operation in INTT, the acceleration scheme designs the corresponding computing unit to realize it. At the same time, according to the characteristics of butterfly operation and hardware I/O characteristics, it adopts the method of organizing butterfly operation in the way of serialization connection and carries out parallel design of butterfly operation unit in a small range. In addition, data storage is reorganized to speed up the speed of data access during butterfly operation execution. In order to reduce the consumption of memory and balance the consumption of memory and time, some data used in the process of NTT is stored in advance. And the rest of the data is generated during the execution of NTT operation, so as to ensure the fast implementation speed of the algorithm and less memory consumption. However, run-time-generated data and pre-generated data have their own advantages and disadvantages, which are about the balance between time consumption and space consumption, and should be used according to the actual situation.

For the basic operation of LWE-based FHE, such as RNS, CRT, NTT-based polynomial multiplication, modulo inverse, modulo reduction, and all the other polynomial and scalar operations are carried out to design and implement the hardware acceleration library by Agrawal *et al.* [52]. By mining the parallelism of some basic operations, they adopt the parallelization method for multi-round calculation in RNS, so as to realize the acceleration of the algorithm.

Turan *et al.* [53] (HEAWS) accelerate the BFV based on FPGA, who design multiple parallel coprocessors in the FPGA in order to execute several operations simultaneously. At the same time, they also design an off-chip data transfer strategy to meet the needs of multiple coprocessors. Because the FPGA used by them is cloud-based FPGA, which brings the extra communication overhead brought, the state-of-the-art 512-bit XDMA feature of high bandwidth communication is used to reduce the overhead of HW/SW data transfer. However, one of the innovative points of HEAWS is that it uses cloud-based FPGA for the first time, which improves its availability.

Gener *et al.* [54] accelerate the bootstrapping algorithm of TFHE based on FPGA. They design an accelerator, which adds TFHE-specific custom instruction extensions to an FPGA-based programmable vector engine. This makes linear performance scale as the number of vector lanes increases from 4 to 16. However, this paper only presents a preliminary architecture to accelerate TFHE bootstrapping, which still uses an $o(n^2)$ algorithm of the polynomial multiplier. And as the paper says, this is one direction in which the accelerator designed could be improved.

Fadhli *et al.* [55] mainly realize the systolic arrays by using FPGA, so as to achieve acceleration of the algorithm. Systolic arrays feature the ability to perform multiple calculations on a single input without waiting. At the same time, it can continue to input the required data while calculating.

Ye *et al.* [56] focus on the acceleration of the plaintext multiplication ciphertext operation. In this paper, based on multiple processing elements (PEs), parallelization is designed and realized for Hadamard product and accumulations operations of vectors, which is that the vector is divided into subvectors, and then parallelization operations between subvectors were realized. The communication delay between FPGA and external memory is avoided by means of a double data cache. Besides, based on the designed performance model, the optimal homomorphic parameters are selected for each layer of the neural network, so as to achieve further acceleration of the neural network using the FHE. Compared to prior work, the acceleration scheme first proposes the low latency inference accelerator for convolutional layers of ResNet-50 based on FPGA.

Based on FPGA, a multi-level parallelism degree is explored for FHE by Xin *et al.* [57]. And parallel scheme is designed and implemented to realize algorithm acceleration. This paper mainly implements parallelization acceleration for three levels of the algorithm. NTT parallel implementation is the basic parallel acceleration method. About this acceleration scheme, by designing several butterfly computing units and executing them in parallel, the parallel acceleration of NTT algorithm is realized. RNS parallel implementation is the medium parallel acceleration method. Based on the NTT core composed of multiple butterfly units, it uses three NTT cores to realize acceleration of polynomial multiplication of RNS form. The parallel implementation of two ciphertext operations is the advanced parallel acceleration method. They equip NTT cores for each polynomial in both two ciphertexts and form the high-level parallelism in ciphertexts. Compared with the acceleration scheme [50], under the condition of using the same number of digital signal processor (DSP) blocks, the NTT cores in the acceleration scheme [57] consume fewer resources.

Aiming at the polynomial multiplication operation used in BFV, Syafalni *et al.* [58] propose a method based on a convolution approach to complete the corresponding polynomial multiplication operation. At the same time, this paper also designs a two-dimensional systolic array on the level of hardware implementation, so as to complete the polynomial multiplication with high parallelism. In order to avoid data transmission delay caused by noise generated by software, they first design a corresponding hardware noise generation module.

Based on FPGA, Yang *et al.* [59] design and implement the corresponding acceleration scheme for SCNN which uses FHE. They analyze the memory access requirement in the process of FHE, so as to obtain the memory access pattern. Based on the memory access pattern, they continue to redesign the data flow to avoid memory access conflicts and realize highly parallel execution of the algorithm, so as to complete the acceleration of the algorithm.

Aiming at key-switching operation, which is the performance bottleneck in CKKS, Han *et al.* [60] explore the data dependence in this operation to minimize the interdependence between data. Then, based on this, they maximize the parallel calculation of data and achieve the acceleration of the algorithm.

Agrawal *et al.* [61] (FAB), aiming at the bootstrapping process of large homomorphic parameters, design a new hardware architecture and balance the memory consumption and computing consumption. FAB architecture efficiently utilizes the available U/BRAM blocks on the FPGA as on-chip memory. Mapping the polynomial data bit-width to that of the U/BRAM blocks data width enables storage of up to 43 MB of on-chip data.

Aiming at the bootstrapping of TFHE, the acceleration scheme is designed and implemented by Ye *et al.* [62]. On one hand, to enable efficient multi-level parallelism, they customize the data layout of TFHE ciphertext for FPGA on-chip SRAM to optimize data access and reduce memory access conflict. The data layout also improves data reuse to effectively utilize the external memory bandwidth. On the other hand, the acceleration scheme is parameterized and can be configured to achieve high throughput and low latency for TFHE bootstrapping when different users have specific security requirements. This improves its availability of it in another way compared with [53].

Su *et al.* [63] accelerate NTT based on FPGA. To reduce latency, the acceleration scheme merges the preprocessing and postprocessing into the NTT and INTT, respectively. Besides, a reconfigurable modular multiplier, which is based on DSP, is proposed to speed up the modular multiplication. In order to avoid designing an independent memory access pattern for INTT, a unified read/write structure of NTT/INTT is presented. Furthermore, they propose a novel memory access pattern named "cyclic-sharing" to reduce 25% memory capacity. In summary, the most significant feature of the acceleration scheme is reconfigurability.

In summary, various researchers have proposed FPGA-based hardware acceleration schemes for different FHE algorithms. These schemes focus on optimizing specific operations, such as NTT, polynomial multiplication, and bootstrapping. The key to achieving acceleration is to explore the parallelism of the algorithm and design hardware architectures that can utilize it. Some schemes [48][61][62] also use on-chip memory to store constant factors or pre-generated data to reduce computation time and memory consumption. Cloud-based FPGA is also used to improve the availability of the acceleration scheme [53]. However, there is still room for improvement in some schemes, such as combining homomorphic parameters tuning to further accelerate FHE. TABLE 7 shows the comparison of acceleration schemes based on FPGA. It can be seen that the availability of FHE schemes based on FPGA is relatively high.

TABLE 7
COMPARISON OF ACCELERATION SCHEMES BASED ON FPGA

| Ref | Year | Availability | Data storage optimization | Homomorphic parameters tuning | Accelerated object |
|---|---|---|---|---|---|
| [48] | 2019 | +++ | × | × | BFV |
| [49] | 2019 | +++ | × | × | BFV |
| [50] | 2020 | ++ | × | √ | CKKS |
| [51] | 2020 | ++++ | √ | × | NTT |
| [52] | 2020 | +++ | × | × | Polynomial Multiplication |
| [53] | 2020 | +++ | √ | × | BFV |
| [54] | 2021 | ++ | × | × | Bootstrapping |
| [55] | 2021 | ++ | × | × | BFV |
| [56] | 2021 | +++ | √ | × | BFV |
| [57] | 2021 | ++++ | × | × | CKKS |
| [58] | 2022 | ++ | × | × | Polynomial Multiplication |
| [59] | 2022 | ++ | √ | × | BFV |
| [60] | 2022 | +++ | × | × | Key-switching |
| [61] | 2022 | +++ | √ | × | Bootstrapping |
| [62] | 2022 | ++ | √ | × | Bootstrapping |
| [63] | 2022 | ++ | √ | × | NTT |

## 4.4 ASIC-based

ASIC offers the greatest customization flexibility of any hardware platform. However, this is why it is the most difficult to design accelerated solutions based on ASIC because it requires a lot of expertise. Most of the acceleration schemes based on ASIC aim to optimize the data placement strategy, as well as design the corresponding computational circuit so that it can achieve the best effect in the acceleration of FHE. One weakness based on ASIC is that the production process of ASIC is complex and requires a large number of resources. Therefore, the corresponding ASIC-based acceleration solution is difficult to be practical.

In [64], aiming at the NTT operation used in bootstrapping, the parallelism of it is fully explored. On the basis of the design and integration of multiple PEs, the corresponding memory management is carried out in the algorithm implementation. They design multiple PEs by exploring the degree of parallelism of the algorithm. Each PE can be adjusted dynamically to perform a different task, including NTT, INTT, and point-wise multiplication. Thus, they realize highly parallel algorithm execution based on multiple PEs. Because, in theory, more PEs should lead to more parallel computing. In addition, suitable memory management strategies are also explored for each execution stage of polynomial multiplication by taking into account the number of PEs. Through memory management, they realize the efficient utilization of memory resources, so as to further realize the algorithm acceleration.

Reagen et al. [65] implement the acceleration of FHE applications, named as Cheetah, different from the general FHE algorithm optimization. Instead, Cheetah combines it with the DNN model to optimize the FHE performance. For example, the optimal parameter analysis model is established to select the optimal parameter first. Then the optimal parameters are selected for FHE used by each layer network in the DNN model in order to reduce algorithm complexity. For the dot product operation involved in BFV homomorphic encryption algorithm, the partial alignment operation is carried out. At the same time, the homomorphic operation sequence is optimized for performance-sensitive homomorphic operation, in order to minimize the use of noise budget. A lower noise budget allows smaller homomorphic parameters to be selected. Cheetah also proposes a hardware acceleration architecture, which greatly accelerates the computation speed of the FHE combined DNN model by utilizing the parallel computing resources of hardware devices, so that it can meet the needs of practical applications. Compared with prior work, Cheetah has been further improved in practicability, which makes the application of FHE combined with DNN more in line with the practical demand.

Samardzic et al. [66] (F1) introduce universal programmable hardware accelerators in order to accelerate the BGV based on ASIC. Based on ASIC's characteristics, the accelerator speeds up BGV. For various homomorphic operations performed using wide vectors, such as modular addition, modular multiplication, NTTs (forward and inverse in the same unit), and automorphisms, the tailored functional units (FUs) are designed to accelerate them. Several FUs are also grouped in computational clusters for further acceleration. Besides, the programmable framework proposed in F1 fully explores the memory management of FPGA, and then they propose a multi-level memory management system to accelerate the FHE. F1 also uses decoupled data orchestration to hide main memory latency. At the same time, it implements a single-stage bit-sliced crossbar network [79] that provides full bisection bandwidth. Moreover F1 adopts a static, exposed microarchitecture: all components have fixed latencies, which are exposed to the compiler. Static scheduling simplifies logic throughout the chip. Compared with the acceleration scheme [32], F1 is more available because of its programmability.

Kim et al. [67] (BTS) are optimized for the bootstrapping of CKKS. In view of the problem that different selections of homomorphic parameters will affect the performance of FHE in many ways, they study the time required by different combinations of homomorphic parameters for each slot in the bootstrapping process on the premise of ensuring security, so as to select the optimal homomorphic parameters. As a result, BTS achieves a balance between safety and performance. Besides, a lot of time-consuming functions are analyzed, including NTT, INTT, and BConv. On this basis, two kinds of data parallelism modes are explored, which are residue-polynomial level parallelism (rPLP) and coefficient-level parallelism (CLP). And finally, the CLP method is used to accelerate the parallel algorithm. Based on the CLP, the corresponding parallel execution microarchitecture is designed. At the same time, based on the different types of memory and the data access frequency related to the algorithm, the data storage strategy is optimized to achieve further acceleration of bootstrapping of CKKS. Compared with F1, BTS improves the throughput of bootstrapping.

Kim et al. [68] (ARK) analyze the memory bottleneck for CKKS bootstrapping hardware acceleration process and design the corresponding solution, which solves the problem of on-chip memory limitation for hardware platforms such as CPU, GPU, and FPGA. On one hand, they design the on-the-fly limb extension, which can pre-calculate the data required for PMult and PAdd operation, so as to reduce the number of times they access data to the on-chip memory. On the other hand, they also design specialized FUs for operations for BConv, thus realizing the acceleration of BConv. In addition, a data distribution strategy is set based on access patterns for BConv to speed up data access. The ARK greatly reduces the number of accesses to the memory under the chip, thus greatly speeding up the implementation of the algorithm. However, the corresponding hardware resource consumption also has increased.

Based on ASIC, Samardzic et al. [69] (CraterLake) study the acceleration scheme for the application of FHE combined with DNN. CraterLake mainly focuses on solving the huge computing overhead of key-switching and uses boosted key-switching. The key innovation in boosted key-switching is to expand the input polynomial to use wider coefficients. It reduces a large auxiliary operand for key-switching hint (KSH). CraterLake also designs the CRB unit and the KSHGen unit. The CRB unit exploits the high internal reuse to allow much higher throughput than independent multipliers and adders communicating through the register file. KSHGen implements an optimization approach primarily through hardware, which has been previously implemented in software [87]. In addition, they have optimized scalar modular multipliers and pipeline each multiplier to its energy-optimal point. Compared with F1, CraterLake can support unlimited depth of multiplication. At the same time, compared with HEAX, [49], HEAWS, etc., CraterLake can execute the entire application based on FHE.

Geelen *et al.* [70] (BASALISC) explore ASIC to accelerate bootstrapping. BASALISC is a three-abstraction-layer RISC architecture. In the process of designing PEs for NTT, BASALISC avoids memory access conflicts in order to accelerate the realization of NTT. Meanwhile, for the twiddle factor, BASALISC designs a twiddle factor factory to reduce the number of twiddles needed to be stored. In addition, the multiply-accumulate unit is designed to accelerate the key-switching process of BGV. The result shows that BASALISC achieves a good balance between performance and resource consumption, and a good acceleration effect. Compared with F1, BASALISC provides better security. And compared with HEAX, BASALISC can adapt homomorphic parameters more in line with realistic requirements.

Jiang *et al.* [71] (MATCHA) accelerate TFHE based on ASIC. They first identify the possibility to use approximate integer FFTs and IFFTs to accelerate TFHE without decryption errors. The depth-first iterative conjugate-pair FFT algorithm [89] also is adopted to decrease the computing overhead of a single FFT or IFFT kernel. Moreover, they propose a pipeline flow for MATCHA to support aggressive bootstrapping key unrolling [90][91], which can reduce the number of HAdd, thus achieving further acceleration. In this paper, the acceleration design for TFHE is relatively novel and can achieve better acceleration effect.

Mert *et al.* [72] (Medha) accelerate the homomorphic evaluation for CKKS based on ASIC. Medha is able to flexibly support several homomorphic encryption parameter sets by using a technique, which is called divide-and-conquer. In addition, the corresponding FUs are designed after sufficiently exploring the parallelism of the algorithm implementation. Further, during the design process, the communication efficiency of each unit is taken into account. And a memory-conservative approach is designed to get rid of any off-chip memory access during homomorphic evaluations.

Due to the significant acceleration effect of ASIC-based FHE acceleration solutions, some researchers [65][69] have been studying how to efficiently apply them to DNN to meet practical needs, and have achieved good results. TABLE 8 shows the comparison of acceleration schemes based on ASIC. It can be seen that all schemes for FHE acceleration based on ASIC will optimize data storage policy. which has to do with the accelerated customization flexibility ASIC can provide.

### 4.5 Other Acceleration Schemes

Compared to the hardware mentioned above, a few works have been focused on leveraging other hardware for FHE acceleration, such as process in memory (PIM). PIM is different from all the above hardware. Its biggest feature is that it can be calculated directly in memory. Therefore, the time consumption of data transmission is greatly reduced. Based on this, the algorithm can be accelerated. As a new hardware platform, the design of FHE acceleration scheme based on PIM is more at the theoretical level, and it is also difficult to put into production and application. TABLE 9 shows the comparison of acceleration schemes based on other hardware.

Based on the PIM, which is also called storage and computing fusions, Gupta *et al.* [73] accelerate the key homomorphic operations, including bootstrapping, HAdd, homomorphic subtraction, HMult, polynomial addition, polynomial multiplication, and NTT. The biggest contribution of PIM is avoiding frequent movement of data and speeding up data access. Then the time consumption of data-intensive computing tasks is reduced. Later, they in [74] accelerate GSW and design the pipeline structure based on PIM, so as to accelerate the implementation of the algorithm.

In [75], the analog-digital implementation and circuit implementation of FHE is converted, so that more homomorphic operations can be supported. At the same time, the algorithm implementation is accelerated.

### 4.6 Hardware Acceleration Summary

Compared with acceleration based on algorithm optimization, scholars pay more attention to hardware-based FHE acceleration. Specifically, a lot of researchers focus on it based on FPGA. This is because, on the one hand, FPGA can provide parallelism similar to GPU, while at the same time, they can provide customization similar to ASIC. Therefore, FPGA has the greatest practicability, which can realize the ideal acceleration effect and facilitate more extensive applications. Compared with FPGA, GPU-based acceleration schemes are easier to use and still have good acceleration effects. Moreover, the acceleration of FHE based on GPU is gradually mature, forming some GPU acceleration libraries that can be directly called. There are two main types of hardware-based acceleration. One is to make full use of the parallel computing resources of the hardware platform, by fully exploring the parallelism of the algorithm, especially for NTT, and then design the pipeline structure. This kind of optimization method can fully combine the parallel computing resources of hardware with the parallel execution of the algorithm, so as to accelerate FHE. The other is that, based on hardware memory characteristics and algorithmic data access patterns, different data generation and storage strategies are designed, such as data precomputation, data runtime generation, data preplacement, and other operations. This kind of optimization method can speed up the data access speed and thus accelerate the implementation of the algorithm. The advantage of a hardware-based acceleration solution is that it can achieve a better acceleration effect, while it requires hardware support and is difficult to develop, especially for hardware platforms such as FPGA and ASIC.

TABLE 8
COMPARISON OF ACCELERATION SCHEMES BASED ON ASIC

| Ref | Year | Availability | Optimize data storage policy | Homomorphic parameters tuning | Accelerated object |
|---|---|---|---|---|---|
| [64] | 2021 | ++ | √ | × | Bootstrapping |
| [65] | 2021 | ++ | √ | √ | BFV |
| [66] | 2021 | ++++ | √ | × | BGV |
| [67] | 2022 | +++ | √ | √ | Bootstrapping |
| [68] | 2022 | +++ | √ | × | Bootstrapping |
| [69] | 2022 | ++++ | √ | × | CKKS |
| [70] | 2022 | +++ | √ | × | BGV |
| [71] | 2022 | +++ | √ | × | TFHE |
| [72] | 2022 | +++ | √ | × | CKKS |

TABLE 9
COMPARISON OF ACCELERATION SCHEMES BASED ON OTHER HARDWARE

| Ref | Year | Availability | Optimize data storage policy | Homomorphic parameters tuning | Accelerated object |
|---|---|---|---|---|---|
| [73] | 2021 | ++ | √ | × | Bootstrapping |
| [74] | 2022 | +++ | √ | × | GSW |
| [75] | 2022 | ++ | √ | √ | Bootstrapping |

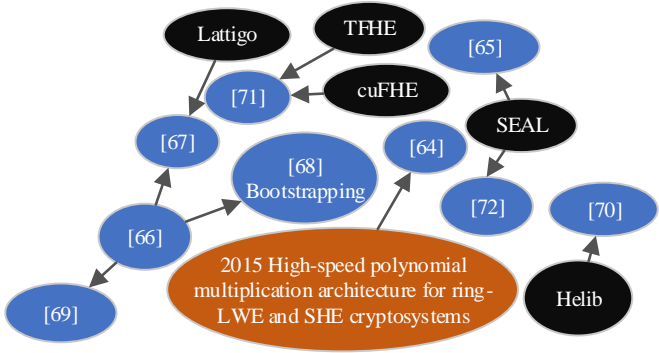

Figure 4. Reference relationship for comparison of acceleration effects of ASIC-based FHE acceleration schemes.

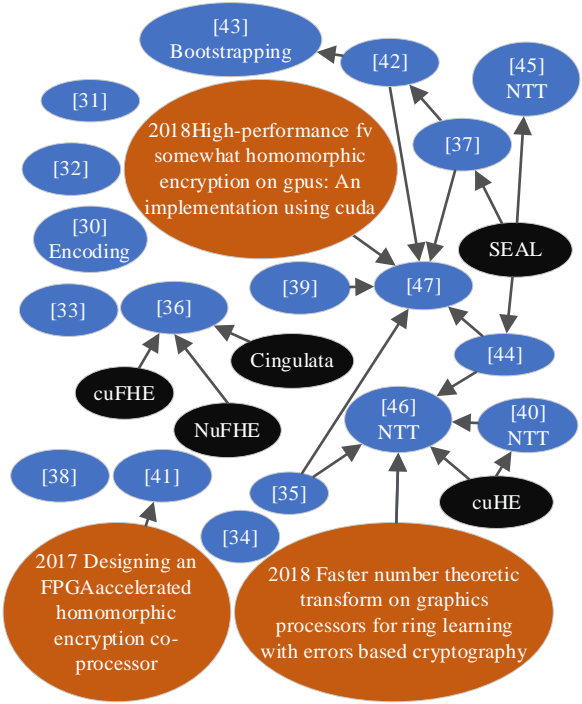

Figure 5. Reference relationship for comparison of acceleration effects of GPU-based FHE acceleration schemes.

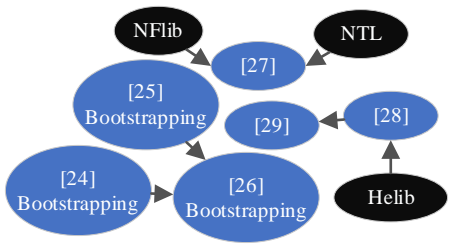

Figure 6. Reference relationship for comparison of acceleration effects of CPU-based FHE acceleration schemes.

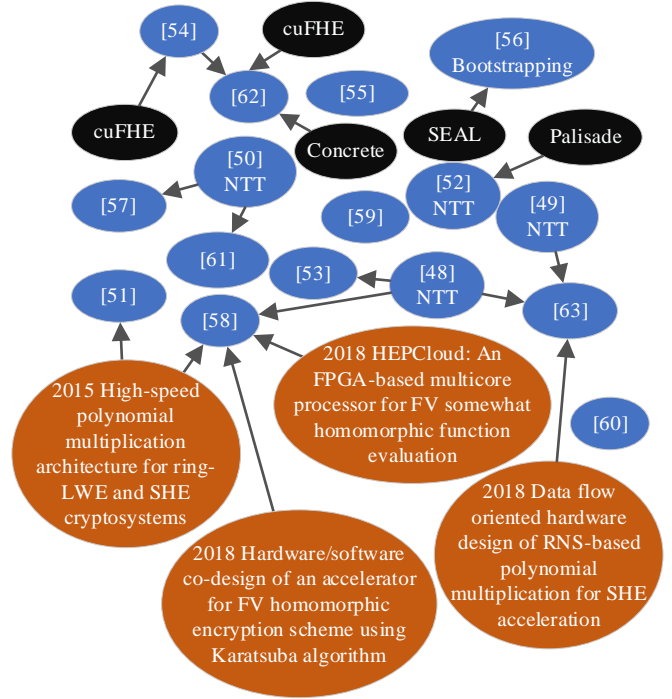

Figure 7. Reference relationship for comparison of acceleration effects of FPGA-based FHE acceleration schemes.

## 5 Challenges and Future Research Directions

In this paper, we review FHE acceleration from two aspects: algorithm-based acceleration and hardware-based acceleration. Owing to the considerable potential for privacy protection provided by FHE while the bottleneck lies in its performance, a rapid development is witnessed in accelerating FHE. The chosen of hardware used for acceleration is from the common CPU to the complex ASIC, which is more suitable for FHE acceleration, with more and more algorithmic optimizations applied. Despite the progress made in accelerating FHE, there remain several open problems, and the performance of FHE is yet to mature to meet the demands of real-time services. To inspire the following research of it, here we list several future directions worthy of further investigation.

**Homomorphic parameters selection.** The selection of homomorphic parameters also needs further research while a few studies on it. The homomorphic parameters selection not only decides the security of FHE, but also a reasonable selection of it can further improve the performance of FHE. It is quite a complex process, while each scheme has specifically selected parameters, all of which are interlinked. And these parameters are usually selected based on current possible lattice-based attacks and their existing limits. An example of a weakness in this approach to parameter selection was exploited in an attack by Lee [100], where the parameter selection in the scheme [101] was not conservative enough to prevent a lattice-based attack exploiting the sparse subset sum problem. In addition, according to different application requirements, there are usually different optimal homomorphic parameters. For example, when FHE is applied to DNN, the corresponding optimal parameter can be selected for each layer of the network. Therefore, more efforts into parameter selection are needed, which must ensure the most suitable parameters are chosen to guarantee both efficiency and security.

**Memory management.** Memory storage is another major bottleneck in the implementation of practical FHE. It involves the use of large parameter sizes and very large ciphertext sizes, which can consume significant amounts of memory. As a result, memory management becomes crucial in FHE implementations. Different FHE schemes, different execution stages of the same FHE scheme, will have different data access patterns. At the same time, different hardware has corresponding memory structures. Existing work explores the data access pattern and data dependency, and designs memory management strategies, including data pregeneration, data preplacement, and data reuse, which also take the characteristics of hardware into account. For example, some methods optimize the utilization of highly accessed data by placing it in the fastest accessible memory in the hardware. However, the current research on memory management strategy is still not enough. Taking GPU as an example, the programming flexibility it provides limits the design of optimal memory management strategies. In addition, most of the research on memory management focuses on NTT and bootstrapping. Therefore, the research on the design of better memory management strategies based on hardware that can provide more flexible memory management needs to be further developed. How to combine hardware characteristics with specific algorithms to minimize storage overhead is important. Especially for special devices, good memory management is extremely important. At the same time, it is necessary to further explore the data dependency and data access pattern of other FHE primitive operations.

**More suitable hardware.** At present, the GPU, FPGA, and ASIC are the major hardware used for FHE acceleration. Among them, the research on GPU has been relatively mature and there are some GPU-based FHE acceleration libraries, including cuHE, cuFHE [102]. In practice, however, the GPU is not suitable for FHE acceleration because of its bandwidth bottleneck. This is because when GPU performs a computing task, data must be transferred from CPU to GPU and then sent back to CPU after the task execution. The large amount of data that FHE generates results in a large memory bandwidth requirement. FPGA and ASIC also have the problem of memory limitation. In addition, due to the specificity of AISC itself, which means that manufacturing ASIC is costly and difficult for researchers to manufacture ASIC, the acceleration scheme based on it only stays in emulation. And they are difficult to directly evaluate and apply, although acceleration schemes based on it have the best acceleration results. Therefore, it is necessary for further research to use new hardware to accelerate FHE. For example, Gupta *et al.* [73][74] try to use PIM to accelerate FHE. At one end of the spectrum, it is still of great research value to design better FHE acceleration schemes based on existing hardware.

**Application convenience and flexibility.** There are many acceleration schemes for FHE, but a few of them are easy to use. The construction theory of FHE is much more complex than the basic symmetric and asymmetric encryption algorithms, such as AES, DES, and RSA, which undoubtedly makes it more difficult for non-professionals to use it. Moreover, since almost all existing acceleration schemes rely on hardware, extra hardware knowledge hinders the use of FHE acceleration schemes even for cryptography professionals. Therefore, the convenience of the accelerated program needs to be further improved. At the same time, different application scenarios have different application requirements for FHE. Therefore, a general acceleration framework also requires further investigation, with the ability to meet different application requirements.

**Algorithm optimization.** At present, most research works on using algorithm optimization to accelerate FHE focus on just applying existing algorithms. Unfortunately, this does not address the inherent performance bottleneck of the FHE itself. Especially for the bootstrapping process, due to its huge computing resources and storage overhead, FHE is often applied in the practical application process in the way of leveled FHE. Therefore, if we want to significantly improve the performance of FHE, we need to make a theoretical breakthrough in FHE itself, and this is a very worthy direction for further research, although it is extremely difficult. Still, further research is needed to explore existing algorithms and techniques that are more suitable for FHE acceleration. For example, batching techniques proposed for FHE schemes could greatly improve the performance of any implementation and should also be investigated further.

In summary, while theoretical innovations in FHE hold promise for greatly improving performance, it may take a significant amount of time to realize these improvements. In the short term, the hardware-based acceleration of FHE is bringing us closer to practical FHE. In addition, how to effectively combine various optimization methods to achieve fine-grained FHE acceleration is also an issue that researchers need to consider. Figure 4-7 respectively shows the reference relationship for comparison of acceleration effects of FHE acceleration schemes based on different hardware. The articles in blue represent themselves discussed in this paper, which are related to FHE acceleration and published between 2019 and 2022. The articles in orange are related to FHE acceleration, but were published earlier than 2019. The black node represents the FHE implementation library. In the process of drawing them, we found that there is no good baseline when comparing the acceleration effect of FHE acceleration schemes. Therefore, how to create a reasonable baseline is also a valuable research direction, which can promote the development of FHE accelerated research.

## 6 Conclusion

By conducting a thorough analysis and comparison of various methods for accelerating fully homomorphic encryption (FHE), we comprehensively explored the future research directions of homomorphic encryption acceleration from multiple perspectives. The primary objective of this article is to provide researchers in the field of homomorphic encryption acceleration with a clear, comprehensive, and in-depth perspective, in order to facilitate a better understanding and application of homomorphic encryption technology, and to offer valuable guidance and support for the development of homomorphic encryption in practical applications and theoretical research. It is our belief that the novel ideas and solutions presented in this article will have a positive impact and drive on the research and application of homomorphic encryption acceleration, and promote the advancement and dissemination of FHE.

## References

[1] Jin Li, Heng Ye, Tong Li, Wei Wang, Wenjing Lou, Y. Thomas Hou, Jiqiang Liu, Rongxing Lu: Efficient and Secure Outsourcing of Differentially Private Data Publishing With Multiple Evaluators. IEEE Trans. Dependable Secur. Comput. 19(1): 67-76 (2022)
[2] Payam Hanafizadeh, Ahad Zare Ravasan: A Systematic Literature Review on IT Outsourcing Decision and Future Research Directions. J. Glob. Inf. Manag. 28(2): 160-201 (2020)
[3] Joppe W. Bos, Kristin E. Lauter, Michael Naehrig: Private predictive analysis on encrypted medical data. J. Biomed. Informatics 50: 234-243 (2014)
[4] Jiayi Hu, Jiahui Deng, Wenqing Wan, Jiawei Qian: Multi-party Secure Computing Financial Shared Platform Based on Lightweight Privacy Protection under FHE. 2020 International Conference on Artificial Intelligence and Computer Engineering. IEEE, 2020: 245-249
[5] Rini Deviani, Sri Azizah Nazhifah, Aulia Syarif Aziz: Fully Homomorphic Encryption for Cloud Based E-Government Data. Cyberspace: Jurnal Pendidikan Teknologi Informasi 2022: 105-118
[6] Marten van Dijk, Ari Juels: On the Impossibility of Cryptography Alone for Privacy-Preserving Cloud Computing. HotSec 2010
[7] Wilson Abel Alberto Torres, Nandita Bhattacharjee, Bala Srinivasan: Effectiveness of Fully Homomorphic Encryption to Preserve the Privacy of Biometric Data. iiWAS 2014: 152-158
[8] Ronald L. Rivest, Len Adleman, Michael L. Dertouzos: On data banks and privacy homomorphisms. Foundations of secure computation, 1978, 4(11): 169-180
[9] Taher El Gamal: A public key cryptosystem and a signature scheme based on discrete logarithms. IEEE Trans. Inf. Theory 31(4): 469-472 (1985)
[10] Pascal Paillier: Public-Key Cryptosystems Based on Composite Degree Residuosity Classes. EUROCRYPT 1999: 223-238
[11] Andrew Chi-Chih Yao: Protocols for Secure Computations (Extended Abstract). FOCS 1982: 160-164
[12] Tomas Sander, Adam L. Young, Moti Yung: Non-Interactive CryptoComputing For NC1. FOCS 1999: 554-567
[13] Dan Boneh, Eu-Jin Goh, Kobbi Nissim: Evaluating 2-DNF Formulas on Ciphertexts. TCC 2005: 325-341
[14] Craig Gentry: A fully homomorphic encryption scheme[M]. Stanford university, 2009.
[15] Zvika Brakerski, Craig Gentry, Vinod Vaikuntanathan: (Leveled) Fully Homomorphic Encryption without Bootstrapping. ACM Trans. Comput. Theory 6(3): 13:1-13:36 (2014)
[16] Junfeng Fan, Frederik Vercauteren: Somewhat Practical Fully Homomorphic Encryption. IACR Cryptol. ePrint Arch. 2012: 144 (2012)
[17] Craig Gentry, Amit Sahai, Brent Waters: Homomorphic Encryption from Learning with Errors: Conceptually-Simpler, Asymptotically-Faster, Attribute-Based. CRYPTO (1) 2013: 75-92
[18] Jung Hee Cheon, Andrey Kim, Miran Kim, Yong Soo Song: Homomorphic Encryption for Arithmetic of Approximate Numbers. ASIACRYPT (1) 2017: 409-437
[19] Ciara Moore, Máire O'Neill, Elizabeth O'Sullivan, Yarkin Doröz, Berk Sunar: Practical homomorphic encryption: A survey. ISCAS 2014: 2792-2795
[20] Abbas Acar, Hidayet Aksu, A. Selcuk Uluagac, Mauro Conti: A Survey on Homomorphic Encryption Schemes: Theory and Implementation. ACM Comput. Surv. 51(4): 79:1-79:35 (2018)
[21] Bechir Alaya, Lamri Laouamer, Nihel Msilini: Homomorphic encryption systems statement: Trends and challenges. Comput. Sci. Rev. 36: 100235 (2020)
[22] Alexander Wood, Kayvan Najarian, Delaram Kahrobaei: Homomorphic Encryption for Machine Learning in Medicine and Bioinformatics. ACM Comput. Surv. 53(4): 70:1-70:35 (2021)
[23] Chiara Marcolla, Victor Sucasas, Marc Manzano, Riccardo Bassoli, Frank H. P. Fitzek, Najwa Aaraj: Survey on Fully Homomorphic Encryption, Theory, and Applications. Proc. IEEE 110(8): 1572-1609 (2022)
[24] Hao Chen, Ilaria Chillotti, Yongsoo Song: Improved Bootstrapping for Approximate Homomorphic Encryption. EUROCRYPT (2) 2019: 34-54
[25] Kyoohyung Han, Dohyeong Ki: Better Bootstrapping for Approximate Homomorphic Encryption. CT-RSA 2020: 364-390
[26] Jean-Philippe Bossuat, Christian Mouchet, Juan Ramón Troncoso-Pastoriza, Jean-Pierre Hubaux: Efficient Bootstrapping for Approximate Homomorphic Encryption with Non-sparse Keys. EUROCRYPT (1) 2021: 587-61
[27] Fabian Boemer, Sejun Kim, Gelila Seifu, Fillipe D. M. de Souza, Vinodh Gopal: Intel HEXL: Accelerating Homomorphic Encryption with Intel AVX512-IFMA52. WAHC@CCS 2021: 57-62
[28] Yu Ishimaki, Hayato Yamana: Faster Homomorphic Trace-Type Function Evaluation. IEEE Access 9: 53061-53077 (2021)
[29] Kotaro Inoue, Takuya Suzuki, Hayato Yamana: Acceleration of Homomorphic Unrolled Trace-Type Function using AVX512 instructions. WAHC@CCS 2022: 47-52
[30] Chao Jin, Ahmad Al Badawi, Balagopal Unnikrishnan, Jie Lin, Chan Fook Mun, James M. Brown, J. Peter Campbell, Michael Chiang, Jayashree Kalpathy-Cramer, Vijay Ramaseshan Chandrasekhar, Pavitra Krishnaswamy, Khin Mi Mi Aung: CareNets: Efficient homomorphic CNN for high resolution images. NeurIPS Workshop on Privacy in Machine Learning (PriML). 2019
[31] Ahmad Al Badawi, Yuriy Polyakov, Khin Mi Mi Aung, Bharadwaj Veeravalli, Kurt Rohloff: Implementation and performance evaluation of RNS variants of the BFV homomorphic encryption scheme. IEEE Transactions on Emerging Topics in Computing, 2019, 9(2): 941-956
[32] Cristian Lupascu, Mihai Togan, Victor Valeriu Patriciu: Acceleration Techniques for Fully-Homomorphic Encryption Schemes. CSCS 2019: 118-122
[33] Xinya Lei, Ruixin Guo, Feng Zhang, Lizhe Wang, Rui Xu, Guangzhi Qu: Accelerating Homomorphic Full Adder Based on FHEW Using Multicore CPU and GPUs. HPCC/SmartCity/DSS 2019: 2508-2513
[34] Jing Xia, Zhong Ma, Xinfa Dai: Parallel Computing Mode in Homomorphic Encryption Using GPUs Acceleration in Cloud. J. Comput. 14(7): 451-469 (2019)
[35] Sangpyo Kim, Wonkyung Jung, Jaiyoung Park, Jung Ho Ahn: Accelerating Number Theoretic Transformations for Bootstrappable Homomorphic Encryption on GPUs. IISWC 2020: 264-275
[36] Toufique Morshed, Md Momin Al Aziz, Noman Mohammed: CPU and GPU Accelerated Fully Homomorphic Encryption. HOST 2020: 142-153
[37] Ahmad Al Badawi, Louie Hoang, Chan Fook Mun, Kim Laine, Khin Mi Mi Aung: PrivFT: Private and Fast Text Classification With Homomorphic Encryption. IEEE Access 8: 226544-226556 (2020)
[38] Ahmad Al Badawi, Bharadwaj Veeravalli, Jie Lin, Xiao Nan, Kazuaki Matsumura, Khin Mi Mi Aung: Multi-GPU Design and Performance Evaluation of Homomorphic Encryption on GPU Clusters. IEEE Trans. Parallel Distributed Syst. 32(2): 379-391 (2021)
[39] Pedro Geraldo M. R. Alves, Jheyne N. Ortiz, Diego F. Aranha: Faster Homomorphic Encryption over GPGPUs via Hierarchical DGT. Financial Cryptography (2) 2021: 520-540
[40] Jia-Zheng Goey, Wai-Kong Lee, Bok-Min Goi, Wun-She Yap: Accelerating number theoretic transform in GPU platform for fully homomorphic encryption. J. Supercomput. 77(2): 1455-1474 (2021)
[41] Wonkyung Jung, Eojin Lee, Sangpyo Kim, Jongmin Kim, Namhoon Kim, Keewoo Lee, Chohong Min, Jung Hee Cheon, Jung Ho Ahn: Accelerating Fully Homomorphic Encryption Through Architecture-Centric Analysis and Optimization. IEEE Access 9: 98772-98789 (2021)
[42] Wonkyung Jung, Sangpyo Kim, Jung Ho Ahn, Jung Hee Cheon, Younho Lee: Over 100x Faster Bootstrapping in Fully Homomorphic Encryption through Memory-centric Optimization with GPUs. IACR Trans. Cryptogr. Hardw. Embed. Syst. 2021(4): 114-148 (2021)
[43] Leo de Castro, Rashmi Agrawal, Rabia Tugce Yazicigil, Anantha P. Chandrakasan, Vinod Vaikuntanathan, Chiraag Juvekar, Ajay Joshi: Does Fully Homomorphic Encryption Need Compute Acceleration? CoRR abs/2112.06396 (2021)
[44] Özgün Özerk, Can Elgezen, Ahmet Can Mert, Erdinç Öztürk, Erkay Savas: Efficient number theoretic transform implementation on GPU for homomorphic encryption. J. Supercomput. 78(2): 2840-2872 (2022)
[45] Enes Recep Türkoglu, Ali Sah Özcan, Can Ayduman, Ahmet Can Mert, Erdinç Öztürk, Erkay Savas: An Accelerated GPU Library for Homomorphic Encryption Operations of BFV Scheme. ISCAS 2022: 1155-1159
[46] Kaustubh Shivdikar, Gilbert Jonatan, Evelio Mora, Neal Livesay, Rashmi Agrawal, Ajay Joshi, José L. Abellán, John Kim, David R. Kaeli: Accelerating Polynomial Multiplication for Homomorphic Encryption on GPUs. SEED 2022: 61-72

[47] Shiyu Shen, Hao Yang, Yu Liu, Zhe Liu, Yulei Zhao: CARM: CUDA-Accelerated RNS Multiplication in Word-Wise Homomorphic Encryption Schemes for Internet of Things. IEEE Transactions on Computers, 2022

[48] Sujoy Sinha Roy, Furkan Turan, Kimmo Järvinen, Frederik Vercauteren, Ingrid Verbauwhede: FPGA-Based High-Performance Parallel Architecture for Homomorphic Computing on Encrypted Data. HPCA 2019: 387-398

[49] Ahmet Can Mert, Erdinç Öztürk, Erkay Savas: Design and Implementation of Encryption/Decryption Architectures for BFV Homomorphic Encryption Scheme. IEEE Trans. Very Large Scale Integr. Syst. 28(2): 353-362 (2020)

[50] M. Sadegh Riazi, Kim Laine, Blake Pelton, Wei Dai: HEAX: An Architecture for Computing on Encrypted Data. ASPLOS 2020: 1295-1309

[51] Sunwoong Kim, Keewoo Lee, Wonhee Cho, Yujin Nam, Jung Hee Cheon, Rob A. Rutenbar: Hardware Architecture of a Number Theoretic Transform for a Bootstrappable RNS-based Homomorphic Encryption Scheme. FCCM 2020: 56-64

[52] Rashmi S. Agrawal, Lake Bu, Michel A. Kinsy: Fast Arithmetic Hardware Library For RLWE-Based Homomorphic Encryption. FCCM 2020: 206

[53] Furkan Turan, Sujoy Sinha Roy, Ingrid Verbauwhede: HEAWS: An Accelerator for Homomorphic Encryption on the Amazon AWS FPGA. IEEE Trans. Computers 69(8): 1185-1196 (2020)

[54] Gener Serhan, Newton Parker, Tan Daniel, Richelson Silas, Lemieux, Guy, Brisk Philip: An fpga-based programmable vector engine for fast fully homomorphic encryption over the torus. SPSL: Secure and Private Systems for Machine Learning (ISCA Workshop). 2021

[55] Hamdani Fadhli, Infall Syafalni, Nana Sutisna, Rahmat Mulyawan, M. Iqbal Arsyad, Trio Adiono: Accelerating Homomorphic Encryption using Systolic Arrays with Polynomial Optimization. 2021 International Symposium on Electronics and Smart Devices (ISESD). IEEE, 2021: 1-6

[56] Tian Ye, Sanmukh R. Kuppannagari, Rajgopal Kannan, Viktor K. Prasanna: Performance Modeling and FPGA Acceleration of Homomorphic Encrypted Convolution. FPL 2021: 115-121

[57] Guozhu Xin, Yifan Zhao, Jun Han: A Multi-Layer Parallel Hardware Architecture for Homomorphic Computation in Machine Learning. ISCAS 2021: 1-5

[58] Infall Syafalni, Gilbert Jonatan, Nana Sutisna, Rahmat Mulyawan, Trio Adiono: Efficient Homomorphic Encryption Accelerator With Integrated PRNG Using Low-Cost FPGA. IEEE Access 10: 7753-7771 (2022)

[59] Yang Yang, Sanmukh R. Kuppannagari, Rajgopal Kannan, Viktor K. Prasanna: FPGA Accelerator for Homomorphic Encrypted Sparse Convolutional Neural Network Inference. FCCM 2022: 1-9

[60] Mingqin Han, Yilan Zhu, Qian Lou, Zimeng Zhou, Shanqing Guo, Lei Ju: coxHE: A software-hardware co-design framework for FPGA acceleration of homomorphic computation. DATE 2022: 1353-1358

[61] Rashmi Agrawal, Leo de Castro, Guowei Yang, Chiraag Juvekar, Rabia Tugce Yazicigil, Anantha P. Chandrakasan, Vinod Vaikuntanathan, Ajay Joshi: FAB: An FPGA-based Accelerator for Bootstrappable Fully Homomorphic Encryption. CoRR abs/2207.11872 (2022)

[62] Tian Ye, Rajgopal Kannan, Viktor K. Prasanna: FPGA Acceleration of Fully Homomorphic Encryption over the Torus. HPEC 2022: 1-7

[63] Yang Su, Bai-Long Yang, Chen Yang, Zepeng Yang, Yi-Wei Liu: A Highly Unified Reconfigurable Multicore Architecture to Speed Up NTT/INTT for Homomorphic Polynomial Multiplication. IEEE Trans. Very Large Scale Integr. Syst. 30(8): 993-1006 (2022)

[64] Weihang Tan, Benjamin M. Case, Gengran Hu, Shuhong Gao, Yingjie Lao: An Ultra-Highly Parallel Polynomial Multiplier for the Bootstrapping Algorithm in a Fully Homomorphic Encryption Scheme. J. Signal Process. Syst. 93(6): 643-656 (2021)

[65] Brandon Reagen, Wooseok Choi, Yeongil Ko, Vincent T. Lee, Hsien-Hsin S. Lee, Gu-Yeon Wei, David Brooks: Cheetah: Optimizing and Accelerating Homomorphic Encryption for Private Inference. HPCA 2021: 26-39

[66] Nikola Samardzic, Axel Feldmann, Aleksandar Krastev, Srinivas Devadas, Ronald G. Dreslinski, Christopher Peikert, Daniel Sánchez: F1: A Fast and Programmable Accelerator for Fully Homomorphic Encryption. MICRO 2021: 238-252

[67] Sangpyo Kim, Jongmin Kim, Michael Jaemin Kim, Wonkyung Jung, John Kim, Minsoo Rhu, Jung Ho Ahn: BTS: an accelerator for bootstrappable fully homomorphic encryption. ISCA 2022: 711-725

[68] Jongmin Kim, Gwangho Lee, Sangpyo Kim, Gina Sohn, Minsoo Rhu, John Kim, Jung Ho Ahn: ARK: Fully Homomorphic Encryption Accelerator with Runtime Data Generation and Inter-Operation Key Reuse. MICRO 2022: 1237-1254

[69] Nikola Samardzic, Axel Feldmann, Aleksandar Krastev, Nathan Manohar, Nicholas Genise, Srinivas Devadas, Karim Eldefrawy, Chris Peikert, Daniel Sánchez: CraterLake: a hardware accelerator for efficient unbounded computation on encrypted data. ISCA 2022: 173-187

[70] Robin Geelen, Michiel Van Beirendonck, Hilder V. L. Pereira, Brian Huffman, Tynan McAuley, Ben Selfridge, Daniel Wagner, Georgios Dimou, Ingrid Verbauwhede, Frederik Vercauteren, David W. Archer: BASALISC: Programmable Asynchronous Hardware Accelerator for BGV Fully Homomorphic Encryption. Cryptology ePrint Archive, 2022

[71] Lei Jiang, Qian Lou, Nrushad Joshi: MATCHA: a fast and energy-efficient accelerator for fully homomorphic encryption over the torus. DAC 2022: 235-240

[72] Ahmet Can Mert, Aikata, Sunmin Kwon, Youngsam Shin, Donghoon Yoo, Yongwoo Lee, Sujoy Sinha Roy: Medha: Microcoded Hardware Accelerator for computing on Encrypted Data. IACR Trans. Cryptogr. Hardw. Embed. Syst. 2023(1): 463-500 (2022)

[73] Saransh Gupta, Tajana Simunic Rosing: Invited: Accelerating Fully Homomorphic Encryption with Processing in Memory. DAC 2021: 1335-1338

[74] Saransh Gupta, Rosario Cammarota, Tajana Rosing: MemFHE: End-to-End Computing with Fully Homomorphic Encryption in Memory. CoRR abs/2204.12557 (2022)

[75] Eduardo Chielle, Oleg Mazonka, Homer Gamil, Michail Maniatakos: Accelerating Fully Homomorphic Encryption by Bridging Modular and Bit-Level Arithmetic. ICCAD 2022: 100:1-100:9

[76] Shai Halevi, Victor Shoup: Faster Homomorphic Linear Transformations in HElib. CRYPTO (1) 2018: 93-120

[77] Sujoy Sinha Roy, Frederik Vercauteren, Nele Mentens, Donald Donglong Chen, Ingrid Verbauwhede: Compact Ring-LWE Cryptoprocessor. CHES 2014: 371-391

[78] Gregor Seiler: Faster AVX2 optimized NTT multiplication for Ring-LWE lattice cryptography. IACR Cryptol. ePrint Arch. 2018: 39 (2018)

[79] Giorgos Passas, Manolis Katevenis, Dionisios N. Pnevmatikatos: Crossbar NoCs Are Scalable Beyond 100 Nodes. IEEE Trans. Comput. Aided Des. Integr. Circuits Syst. 31(4): 573-585 (2012)

[80] Crandall R E: Integer convolution via split-radix fast Galois transform. Center for Advanced Computation Reed College. 1999

[81] https://developer.nvidia.com/cufft

[82] Marten van Dijk, Craig Gentry, Shai Halevi, Vinod Vaikuntanathan: Fully Homomorphic Encryption over the Integers. EUROCRYPT 2010: 24-43

[83] AA Karatsuba, YP Ofman: Multiplication of many-digital numbers by automatic computers. Doklady Akademii Nauk. Russian Academy of Sciences, 1962, 145(2): 293-294

[84] Jean-Sébastien Coron, Avradip Mandal, David Naccache, Mehdi Tibouchi: Fully Homomorphic Encryption over the Integers with Shorter Public Keys. CRYPTO 2011: 487-504

[85] Arnold Schönhage, Volker Strassen: Schnelle Multiplikation großer Zahlen. Computing 7(3-4): 281-292 (1971)

[86] Wei Dai, Berk Sunar: cuHE: A Homomorphic Encryption Accelerator Library. BalkanCryptSec 2015: 169-186

[87] Shai Halevi, Victor Shoup: Design and implementation of HElib: a homomorphic encryption library. IACR Cryptol. ePrint Arch. 2020: 1481 (2020)

[88] Paul Barrett: Implementing the Rivest Shamir and Adleman Public Key Encryption Algorithm on a Standard Digital Signal Processor. CRYPTO 1986: 311-323

[89] Alexandre Bécoulet, Amandine Verguet: A Depth-First Iterative Algorithm for the Conjugate Pair Fast Fourier Transform. IEEE Trans. Signal Process. 69: 1537-1547 (2021)

[90] Florian Bourse, Michele Minelli, Matthias Minihold, Pascal Paillier: Fast Homomorphic Evaluation of Deep Discretized Neural Networks. CRYPTO (3) 2018: 483-512

[91] Tanping Zhou, Xiaoyuan Yang, Longfei Liu, Wei Zhang, Ningbo Li: Faster Bootstrapping With Multiple Addends. IEEE Access 6: 49868-49876 (2018)

[92] David H. Bailey: FFTs in external of hierarchical memory. SC 1989: 234-24

[93] Shai Halevi, Victor Shoup: Algorithms in HElib. CRYPTO (1) 2014: 554-571

[94] Hao Chen, Wei Dai, Miran Kim, Yongsoo Song: Efficient Homomorphic Conversion Between (Ring) LWE Ciphertexts. ACNS (1) 2021: 460-479

[95] Jean-Claude Bajard, Julien Eynard, M. Anwar Hasan, Vincent Zucca: A Full RNS Variant of FV Like Somewhat Homomorphic Encryption Schemes. SAC 2016: 423-442

[96] Craig Gentry, Shai Halevi, Nigel P. Smart: Homomorphic Evaluation of the AES Circuit. CRYPTO 2012: 850-867

[97] Ilaria Chillotti, Nicolas Gama, Mariya Georgieva, Malika Izabachène: Faster Packed Homomorphic Operations and Efficient Circuit Bootstrapping for TFHE. ASIACRYPT (1) 2017: 377-408

[98] Shai Halevi, Victor Shoup: Bootstrapping for HElib. J. Cryptol. 34(1): 7 (2021)

[99] Jean-Philippe Bossuat, Juan Ramón Troncoso-Pastoriza, Jean-Pierre Hubaux: Bootstrapping for Approximate Homomorphic Encryption with Negligible Failure-Probability by Using Sparse-Secret Encapsulation. ACNS 2022: 521-541

[100] Moon Sung Lee: On the sparse subset sum problem from Gentry-Halevi's implementation of fully homomorphic encryption. IACR Cryptol. ePrint Arch. 2011: 567 (2011)

[101] Craig Gentry, Shai Halevi: Implementing Gentry's Fully-Homomorphic Encryption Scheme. EUROCRYPT 2011: 129-148

[102] 2018. CUDA-accelerated Fully Homomorphic Encryption Library. https://github.com/vernamlab/cuFHE. (2018)